\documentclass[iicol]{sn-jnl}

\usepackage{graphicx}%
\usepackage{multirow}%
\usepackage{amsmath,amssymb,amsfonts}%
\usepackage{amsthm}%
\usepackage{mathrsfs}%
\usepackage[title]{appendix}%
\usepackage{xcolor}%
\usepackage{textcomp}%
\usepackage{manyfoot}%
\usepackage{booktabs}%
\usepackage{algorithm}%
\usepackage{algorithmicx}%
\usepackage{algpseudocode}%
\usepackage{listings}%
\usepackage[separate-uncertainty = true,multi-part-units=single]{siunitx}
\DeclareSIUnit{\molar}{M}

\usepackage[backend=bibtex,style=chem-acs]{biblatex}

\graphicspath{{figures/}}

\theoremstyle{thmstyleone}%
\theoremstyle{thmstyletwo}%

\theoremstyle{thmstylethree}%

\begin{document}


\title[Article Title]{Design and optimization of \textit{in situ} self-functionalizing stress sensors}


\author[1,2,3]{\fnm{Olga} \sur{Vasiljevic}}\email{olga.vasiljevic@sorbonne-universite.fr}
\equalcont{These authors contributed equally to this work.}

\author[1,3]{\fnm{Nicolas} \sur{Harmand}}\email{nicolas.harmand@sorbonne-universite.fr}
\equalcont{These authors contributed equally to this work.}

\author[1,3]{\fnm{Antoine} 
\sur{Hubert}}\email{antoine.hubert@sorbonne-universite.fr}

\author[1,3]{\fnm{Lydia} 
\sur{Kebbal}}\email{lydia.kebbal@etu.sorbonne-universite.fr}

\author[1,3]{\fnm{Volker} 
\sur{Bormuth}}\email{volker.bormuth@sorbonne-universite.fr}

\author[4]{\fnm{Clara} 
\sur{Hayn}}\email{chayn@uni-bonn.de}

\author[2,3]{\fnm{Jonathan} \sur{Fouchard}}\email{jonathan.fouchard@sorbonne-universite.fr}

\author[1,3]{\fnm{Elie} \sur{Wandersman}}\email{elie.wandersman@sorbonne-universite.fr}

\author[2,3]{\fnm{Marie Anne} \sur{Breau}}\email{marie.breau@sorbonne-universite.fr}

\author*[1,3]{\fnm{Lea-Laetitia} \sur{Pontani}}\email{lea-laetitia.pontani@sorbonne-universite.fr}

\affil[1]{\orgdiv{Sorbonne University}, \orgdiv{CNRS}, \orgname{Laboratoire Jean Perrin, LJP}, \orgaddress{F-75005 Paris, France}}

\affil[2]{\orgdiv{Sorbonne University}, \orgdiv{CNRS}, \orgdiv{Inserm}, \orgname{Développement Adaptation et Vieillissement, Dev2A}, \orgaddress{F-75005 Paris, France}}

\affil[3]{\orgdiv{Sorbonne University}, \orgdiv{CNRS}, \orgdiv{Inserm}, \orgname{Institut de Biologie Paris-Seine, IBPS}, \orgaddress{F-75005 Paris, France}}

\affil[4]{\orgdiv{University of Bonn},\orgname{Institute of Reconstructive Neurobiology}, \orgaddress{53127 Bonn, Germany}}


\abstract{Mechanical contributions are crucial regulators of diverse biological processes, yet their \textit{in vivo} measurement remains challenging due to limitations of current techniques, that can be destructive or require complex dedicated setups. 
This study introduces a novel method to synthesize biocompatible, self-functionalizing stress sensors based on inverted emulsions, that can be used to probe stresses inside tissues but can also locally perturb the biological environment through specific binder presentation or drug delivery.
We engineered an optimal design for these inverted emulsions, focusing on finding the balance between the two contradictory constraints: achieving low surface tension for deformability while maintaining emulsion instability for efficient self-functionalization and drug release.
Proof-of-concept experiments in both agarose gels and complex biological systems, including brain organoids and zebrafish embryos, confirm the droplets ability to deform in response to mechanical stress applied within the tissue, to self-functionalize and to release encapsulated molecules locally. 
These versatile sensors offer a method for non-invasive stress measurements and targeted chemical delivery within living biological tissues, giving the potential to overcome current technical barriers in biophysical studies.}



\maketitle

\section{Introduction}\label{sec1}

It is well-established that mechanical forces bear a central role during many biological processes such as morphogenesis~\cite{Heisenberg2013, Lecuit2011, Villedieu2020} or cancer progression~\cite{Kumar2009}. 
At the tissue-scale, these forces can be exerted intrinsically, through the motile and contractile activity of the cells within a tissue, or can be sustained from externally applied forces, for instance through the activity of neighboring tissues or through mechanical coupling via the extracellular matrix~\cite{Stooke-Vaughan2018}. 

A lot of effort has therefore been dedicated in the past decades to map out these forces and the associated mechanical properties of tissues undergoing morphogenetic processes (see~\cite{Villeneuve2025, Sugimura2016} for reviews). 
\textit{In vitro}, various bulk rheometry or tensiometry techniques have been adapted to the use of biological material  \cite{Mary2022, Mazuel2015, Stirbat2013, Mgharbel2009, Foty1994}.  
Micromanipulation assays and microfluidics were also developed to gain insight into the mechanical properties of tissues at different scales \cite{Guevorkian2010, Tlili2022}. 
However, these techniques were mostly applied to cellular aggregates or surface tissues \cite{vonDassow2010, Petridou2021} in living organisms because of the constraints of set-up geometry.

Other tools can be used to infer local mechanical properties and stresses \textit{in vivo}, and at different scales. Among these tools, laser ablation is a powerful technique allowing to map out tension within developing tissues \cite{Hutson2003, Rauzi2008, Breau2017, Porazinski2015}. 
However, its destructive nature imposes to renew the experiment on multiple organisms in order to map tension across space and time. Moreover, an absolute measurement of tension requires to know the material properties of the destructed area. 

{Alternatively, sensors in the form of deformable inclusions - encompassing solid beads and liquid droplets -  have been developed in the past few years, and have been used to measure mechanical stresses \textit{in vivo} with a non-destructive approach (see~\cite{Ding2022} for a review). 
On the one hand, solid beads composed of compressible gels offer several advantages. Their mechanical compressibility can be precisely calibrated under controlled conditions (for instance, by applying osmotic pressure), enabling the measurement of isotropic stresses within biological tissues such as cell aggregates or developing embryos~\cite{Dolega2017, Lee2019, Vian2023}. In addition, these beads can be functionalized to interact specifically with their surrounding environment, making them a versatile tool for probing not only compressive but also tensile stresses~\cite{Mohagheghian2018, Traber2019, Gutierrez2021}.}

{On the other hand, incompressible but deformable sensors allow for the measurement of anisotropic stresses generated inside the tissues~\cite{Souchaud2022, Campas2014, Mongera2018}. 
In particular, oil droplets can be functionalized with binders~\cite{Pontani2012, Lucio2015, Pinon2018, Nagendra2023} in order to induce specific adhesion between the droplets and their environment. In this case, the oil droplets are prepared and functionalized prior to their injection inside the tissues, in the form of an emulsion.
This approach has been applied in explants and cellular aggregates~\cite{Campas2014, Lucio2017}, but remains limited for \textit{in vivo} studies. Indeed, for both solid and liquid sensors, the use of such dispersed solutions (emulsions for liquid sensors, dispersions of beads for solid ones) implies that a mixture of sensors and water has to be injected directly inside biological tissues, which remains challenging for multiple reasons. 
First, the water phase contained in the solution may swell the tissue locally, leading to a large stress and potential tissue damage and exclusion of the sensors from the tissue. Second, the dispersed and inherently heterogeneous nature of the dispersions make it very difficult to inject a limited number of sensors at a precise location. 
Therefore, \textit{in vivo} studies actually turned towards the microinjection of single oil droplets from a pure oil phase~\cite{Mongera2018}, thus loosing the surface functionalization properties of classical biomimetic emulsions but gaining in precision and biocompatibility regarding the injection itself.}

In this paper, we introduce {a new method to synthesize stress measurement sensors based on oil droplets} which can gain secondary functions once injected inside the tissue, such as local drug delivery or surface exposure of proteins. 
This so-called self-functionalization \textit{in situ} is due to the fact that the droplet sensors are made of an inverted emulsion of water droplets in oil. 
The inverted emulsion is designed to be unstable enough so that the inner water droplets progressively fuse at the oil-water outer surface, thus releasing locally any drug that was initially contained in the inner water droplets or exposing any protein grafted on their surface. In order for the inverted emulsion to be unstable, there should therefore be a limited amount of surfactants in the formulation.
{On the other hand, for oil droplets to act as stress sensors, i.e. be deformed in biological tissues, their size has to be at least comparable to the elasto-capillary length $L_{ec}$ which is given by $L_{ec}=\gamma/E$, with $\gamma$ the interfacial tension of the oil droplet and $E$ the young modulus of the surrounding tissue. 
If the condition on the elasto-capillary length is fulfilled, the included oil droplets can be deformed by their environment and these deformations can in turn provide a measurement of local stresses.
Therefore, the smaller the surface tension, the smaller the constraint on the oil droplet size, making it a critical parameter for the formulation of stress sensors. }

{We here present a proof of concept study for the use of such oil droplets as functional stress sensors. 
We describe how to fine tune the balance between self-functionalizing properties - requiring inverted emulsion instability, i.e. a \emph{high enough} surface tension - and stress sensor properties - requiring \emph{low enough} surface tension - so that both functions of the sensors can take place inside biological materials such as soft developing tissues.
We then use a mechanical excitation set-up to validate a mechanical model of such liquid insertions inside model materials.
Finally, since this technique allows one to inject single oil droplets of a chosen size inside the tissue and to precisely localize them, we present a few examples of their incorporation in different materials, from agar gels to zebrafish embryos. }

\section{Results}\label{sec2}

\subsection{Design of the inverted emulsion sensor}

\begin{figure}[hbt!]
\centering
\includegraphics[width=0.477\textwidth]{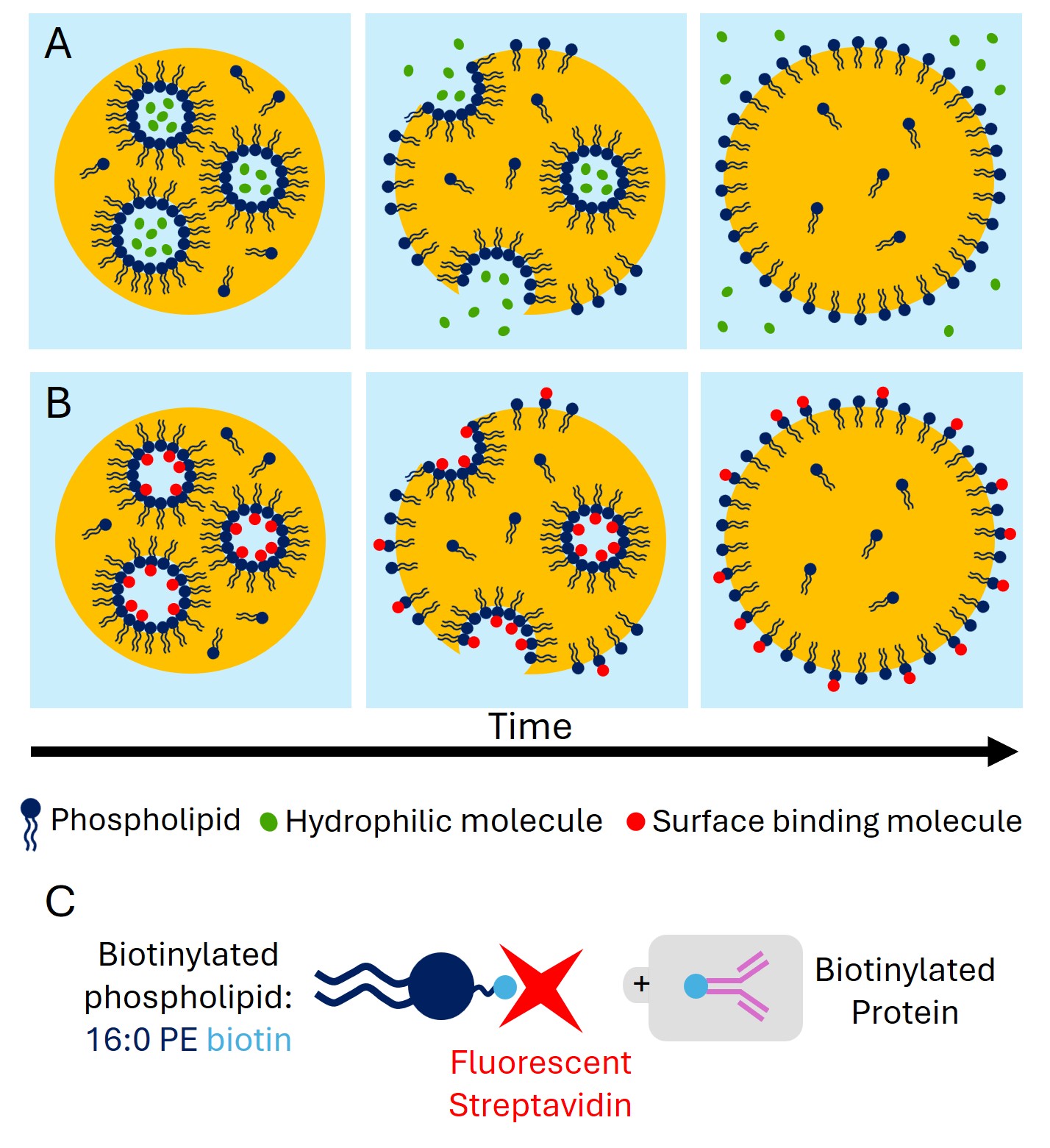}
\caption{ Schematic illustration of an inverted emulsion oil droplet and its different functionalities. (A) Release of hydrophilic molecules from inner water droplets into the surrounding aqueous phase upon destabilization of the inverted emulsion. (B) Self-functionalization of an oil droplet. The surface of the inner water droplets is decorated with molecules through specific interactions with the lipids. Upon destabilization of the inverted emulsion, the inner water droplets fuse with the outer surface of the oil droplet, which results in a redistribution of surface molecules on surface of the oil droplet.
(C) Schematic representation of surface functionalization with phospholipids. Here, the biotinylated phospholipids (16:0 PE biotin) are decorated with fluorescent streptavidin, onto which any biotinylated molecule can be grafted.}\label{fig1}
\end{figure}

In order to combine formulation versatility and ease of the injection process, we explored the idea of using inverted emulsions as a base for the microinjection of isolated oil drops inside tissues. Upon destabilization of the inverted emulsion, the inner water droplets should fuse with the outer oil water interface, thus releasing any chemical that is encapsulated inside them (Fig.\ref{fig1}A). Moreover, if the inner surface of the water droplets is functionalized with proteins or active molecules - that we refer to as binders in the following - the outer surface of the drop can become decorated with the same binders upon fusion (Fig.\ref{fig1}B), thus providing it with self-functionalizing properties. 

In order for this schematic to work, the inverted emulsion has to be unstable. {On the one hand there should therefore not be too much surfactant, which would tend to stabilize the inner water droplets. On the other hand, for the oil droplet to be deformable in biological tissues, there should be enough surfactant in the oil to lower its surface tension.}
In the following, we describe the formation and composition of the inverted emulsion, as well as the systematic measurements of surface tension and destabilization efficiency. These compared measurements allowed us to identify a sweet spot for formulation in which both properties can be fulfilled.

\subsection{Formation of the inverted emulsion}

The inverted emulsion is made of water droplets dispersed in a biocompatible soybean oil containing various concentrations of phospholipids and Tween 20, both of which act as biocompatible surfactants.
The chosen lipid, called 16:0  Biotinyl PE, carries a biotin on its hydrophilic head. This allows us to visualize the surface of the inner water droplets by encapsulating fluorescent streptavidin in the water phase, which gets spontaneously grafted onto the biotinylated lipids heads (Fig.\ref{fig1}C). 
{This lipid was chosen because of its Hydrophilic-Lipophilic Balance index (HLB). This value can be estimated from the size of hydrophilic and hydrophobic domains of a surfactant and provides insight into the structure of a surfactant and its ability to stabilize rather direct (8$<$HLB$<$20) or inverse emulsions (HLB$<$6)~\cite{Griffin1949}. For 16:0  Biotinyl PE, the HLB index yields $\sim$5-6, which is low enough to stabilize water in oil emulsions but still quite hydrophilic. As a point of comparison, a DSPE-PEG(2000) Biotin lipids, that are commonly used to functionalize oil droplets~\cite{Pontani2012}, have an HLB index of $\sim$12-15, which would not be appropriate to populate the surface of the inner water droplets.}

\begin{figure}[hbt!]
\centering
\includegraphics[width=0.477\textwidth]{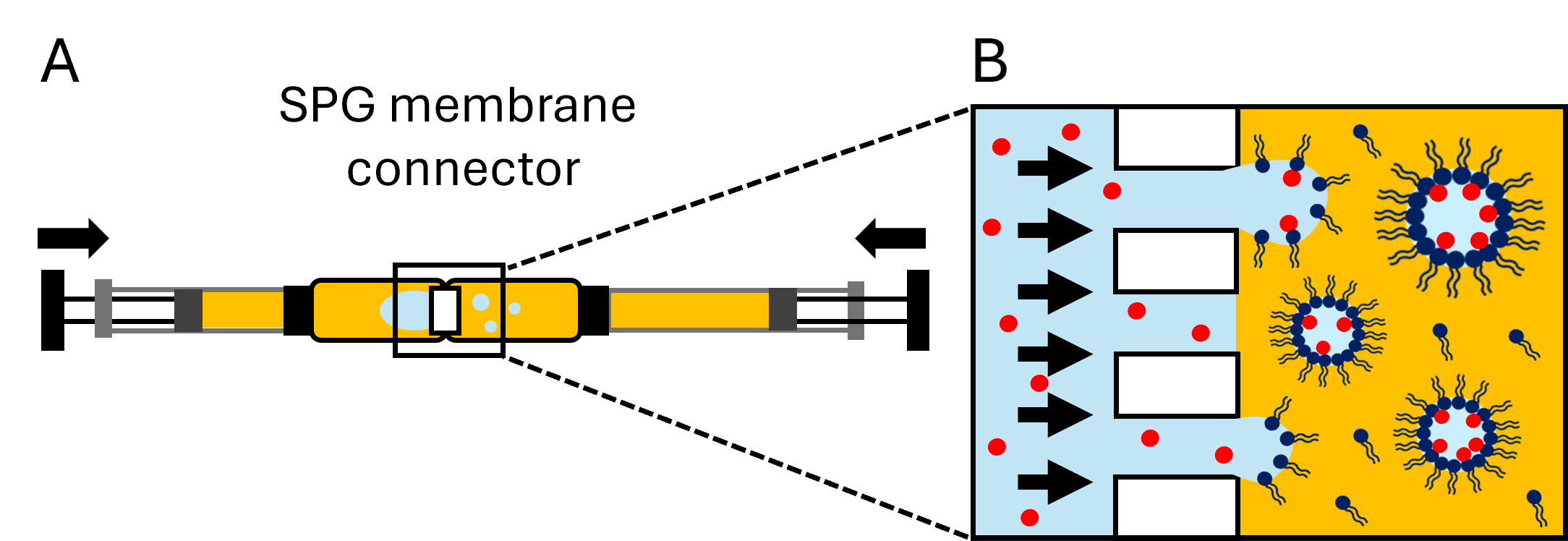}
\caption{Schematic illustration of the emulsification method through the Shirasu Porous Glass (SPG) membrane connector. (A) View of a connector filled with the mix of water and oil phases, with two syringes A and B, locked on both sides. Pushing the syringes back and forth successively results in the repeated extrusion of the aqueous phase through the porous membrane. (B) Scheme of the extrusion at the membrane pore scale. The water phase (light blue) is going through the \SI{5}{\micro\meter} pores of the membrane and the resulting shear creates water droplets in the oil phase (yellow) that are progressively stabilized with the amphiphilic molecules dispersed in the oil phase.}\label{fig2}
\end{figure}

\subsubsection{Emulsification through SPG membranes}

\begin{figure}[hbt!]
\centering
\includegraphics[width=0.477\textwidth]{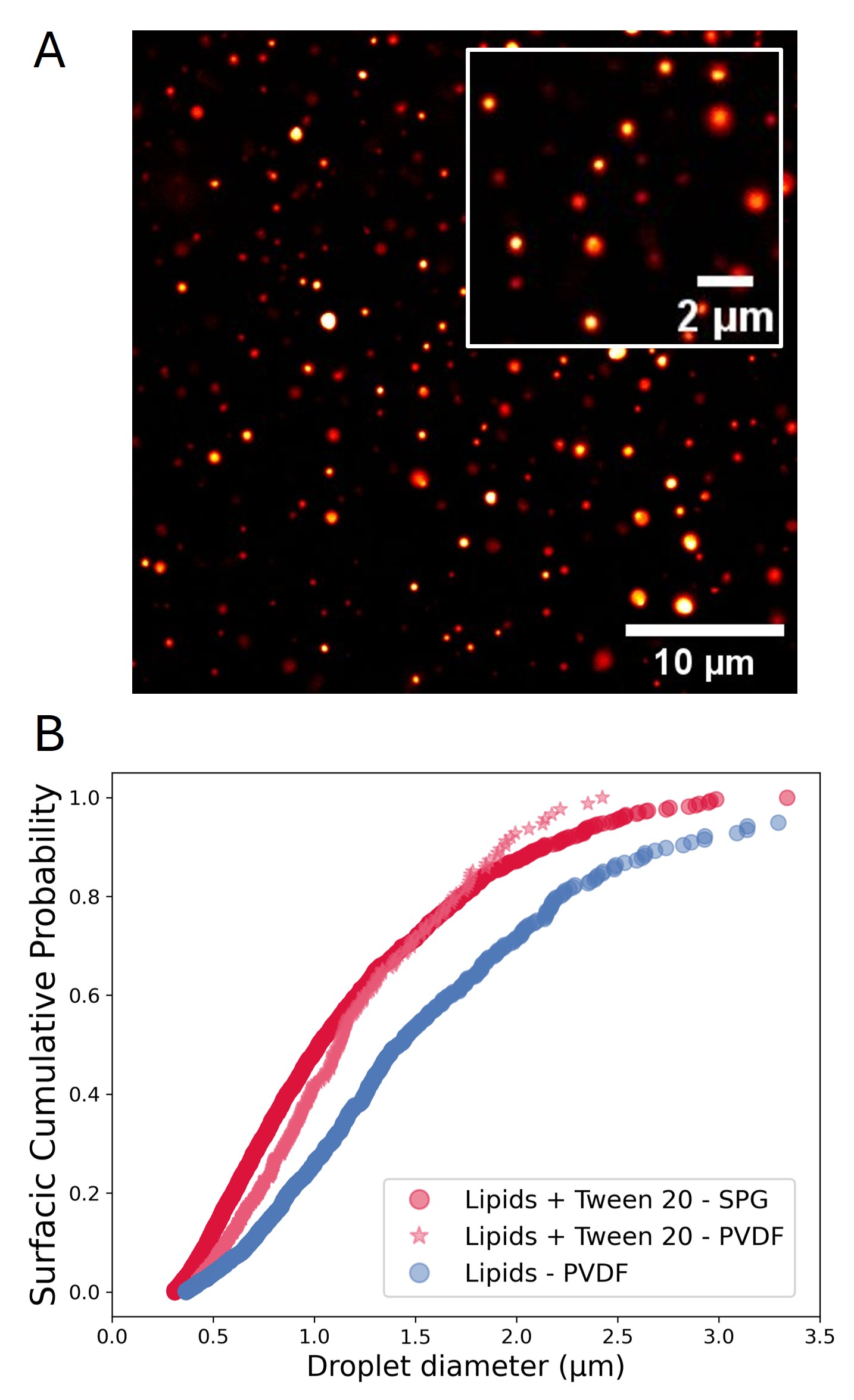}
\caption{Size distribution of the water-in-oil emulsion. (A) Typical field of view of an emulsion made through \SI{5}{\micro\meter} SPG membrane connector. The oil contains \SI{0.5}{\milli\gram\per\milli\liter} lipids and \SI{0.25}{\milli\gram\per\milli\liter} Tween 20. The water droplets are labelled with Streptavidin Alexa 546 and observed through spinning disk confocal microscopy (60X objective). (B) Surfacic Cumulative Distribution Function of water droplet diameters (each droplet has a ponderation in the distribution equal to its surface area) for emulsions prepared with the SPG connector at \SI{0.5}{\milli\gram\per\milli\liter} lipids + \SI{0.25}{\milli\gram\per\milli\liter} Tween 20 (dark pink circles, N=5228 droplets), emulsions using the same formulation but PVDF filters (light pink stars, N=617 droplets), and emulsions with only \SI{0.5}{\milli\gram\per\milli\liter} lipids prepared with PVDF filters (blue circles, N=1296 droplets).
\label{figure_droplet_size}}
\end{figure}

The dispersion of the aqueous phase can be performed with different methods. 
When using expensive and rare biological material (such as antibodies or purified proteins), the emulsification is adapted to the use of small volumes. 
In that case, \SI{50}{\micro\liter} of aqueous phase is gently emulsified in the oil through a porous Shirasu Porous Glass (SPG) membrane (see Fig. \ref{fig2} and Methods). 
This emulsification method only requires very little material (the SPG connector itself and two syringes) and can be performed manually within seconds. 
It is therefore perfectly adapted to the use of fragile biological material and can be implemented straightforwardly in any laboratory environment.
With a membrane of pore size \SI{5}{\micro\meter}, this method typically yields an emulsion with a diameter $\langle d \rangle = 0.66 \pm \SI{0.37}{\micro\meter}$ (mean $\pm$ standard deviation, see Fig.\ref{figure_droplet_size} and Methods).

\subsubsection{Concentrations of lipids and binders}

From their size distributions, we can estimate the quantity of surfactant required to cover the surface area of the inner water droplets in the emulsion. 
The size distribution leads to a mean volume of $\overline{v} = \SI{0.35}{\micro \cubic\meter}$ for the water droplets, while the water volume fraction $\Phi = V_w/ V_{tot}$, with $V_w = \SI{50}{\micro \liter}$ and $V_{tot} = \SI{850}{\micro \litre}$, yields $\Phi = 6.25 \%$ (see Methods). This allows one to calculate the total oil/water interface in the emulsion and the number of lipids needed to cover it assuming a typical surface coverage per lipid of $\sim \SI{1}{\square \nano \meter}$
For the chosen lipids (16:0 Biotinyl PE, $M_p = \SI{940}{\gram \per \mole}$) this leads to a minimum mass concentration of $c_{lipid} \simeq \SI{0.50}{\milli \gram \per \milli \liter}$ to cover the interface of the inner water droplets.
Because the lipids are dispersed in the oil and do not all partition at the oil/water interface, we therefore work at lipid concentrations $c_{lipid} \ \ge \SI{0.50}{\milli \gram \per \milli \liter}$.

The binders concentration also needs to be adapted to the amount of available lipids on the surface. In order to avoid the presence of unattached binders in the inner water droplets,  their concentration must be lower than that of the surface lipids. Moreover, as mentioned above, not all lipids will partition at the oil/water interface, especially in the presence of co-surfactants such as Tween 20. Therefore we work at a ratio of about 1 streptavidin for 2000 lipids. Considering the size of the water droplets in the inverted emulsion, this target value leads to a streptavidin concentration of $c_s = \SI{0.2}{\milli \gram \per \milli \liter}$.

\subsubsection{Emulsification through filters}

If the material contained in the water droplets is more robust and abundant, the emulsification method can be simply performed by pushing the oil and water mix through a classical PVDF (Polyvinylidene fluoride) filter with a \SI{5}{\micro \meter} pore size (see Methods). The dead volume with these filters is quite important, which is why it is not adapted to the handling of precious material such as purified proteins, but can be used for classical fluorophores or drugs.
For the same formulation (see light pink points in Fig. \ref{figure_droplet_size}B), we produced an emulsion that is quite comparable to the one obtained with SPG membranes, with a diameter $\langle d \rangle = 0.79\pm \SI{0.39}{\micro\meter}$ (mean $\pm$ standard deviation).
With an alternative formulation containing only \SI{0.5}{\milli\gram\per\milli\liter} lipids, we get a coarser emulsion, with $\langle d \rangle = 0.90\pm \SI{0.54}{\micro\meter}$ (see blue points in Fig. \ref{figure_droplet_size}B), which is expected since the surface tension of this formulation should be higher (see next section for values).

\subsection{Tuning of the interfacial tension}

\begin{figure}[hbt!]
\centering
\includegraphics[width=0.477\textwidth]{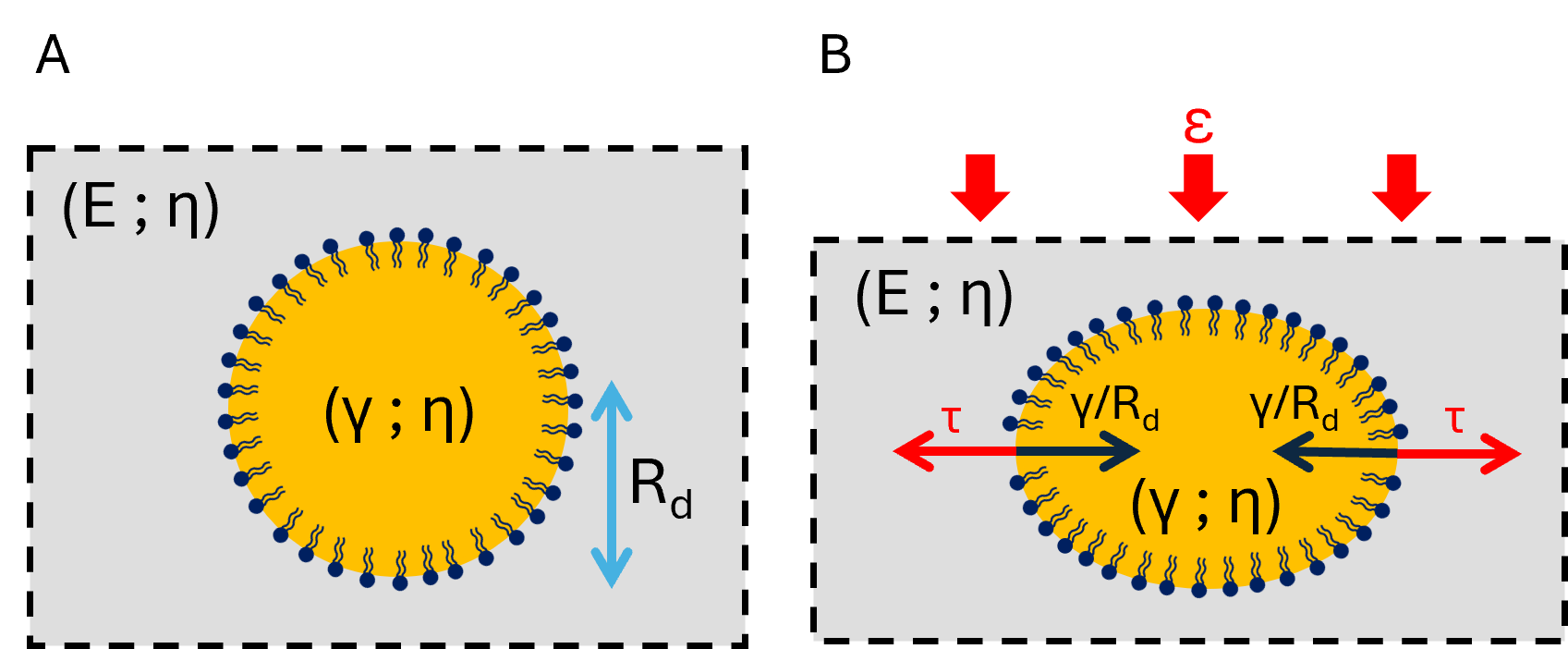}
\caption{Scheme of an oil droplet deformed by a stress imposed on the surrounding medium. (A) Schematic representation of an oil droplet embedded in an elastic medium of Young modulus E. (B) When the medium is under compression, here with an imposed deformation $\epsilon$, it can deform of the oil droplet with a local extensile stress $\tau$, if capillary stresses $\gamma/R_d$ do not dominate.
}\label{deformation_in_compression}
\end{figure}

{In order to use oil droplets as stress sensors in a medium of modulus $E$, their interfacial tension $\gamma$ must be low enough and is therefore a critical parameter of our design.
Indeed, as mentioned in the introduction, if the size $R_d$ of the oil droplet is much smaller than the elastocapillary length $L_{ec}=\gamma/E$, the capillary stresses $\gamma / R_d$ will dominate and the oil droplet will behave as a solid inclusion in the tissue. Conversely, if it is larger or on the order of the elastocapillary length, the local elastic stresses $\tau$ will allow to deform the oil droplet that can then serve as a stress sensor (see Fig.\ref{deformation_in_compression})\cite{Tapie2023}.}

Therefore the radius of the oil droplet $R_d$ should satisfy $R_d\gtrsim \gamma / E$. 
In soft developing tissues or biopolymers such as extracellular matrices, the material Young's modulus typically falls in the \SIrange{100}{1000}{\pascal} range~\cite{bohringerDynamicTractionForce2024}.
The oil droplet size is constrained by the one of the considered biological tissue and the typical length scale of the process being probed. Typically, oil droplets of radius $R_d \sim \SI{10}{\micro\meter}$ are usually small enough to probe cellular scale processes in tissues and large enough to be imaged with a sufficient resolution. 
We therefore use this minimal size to deduce the required maximal interfacial tension for observable oil droplet deformations, which yields:

\[
\gamma \sim E R_d \sim \SIrange{1}{10}{\milli\newton\per\meter}
\]

\begin{figure}[H]
\centering
\includegraphics[width=0.477\textwidth]{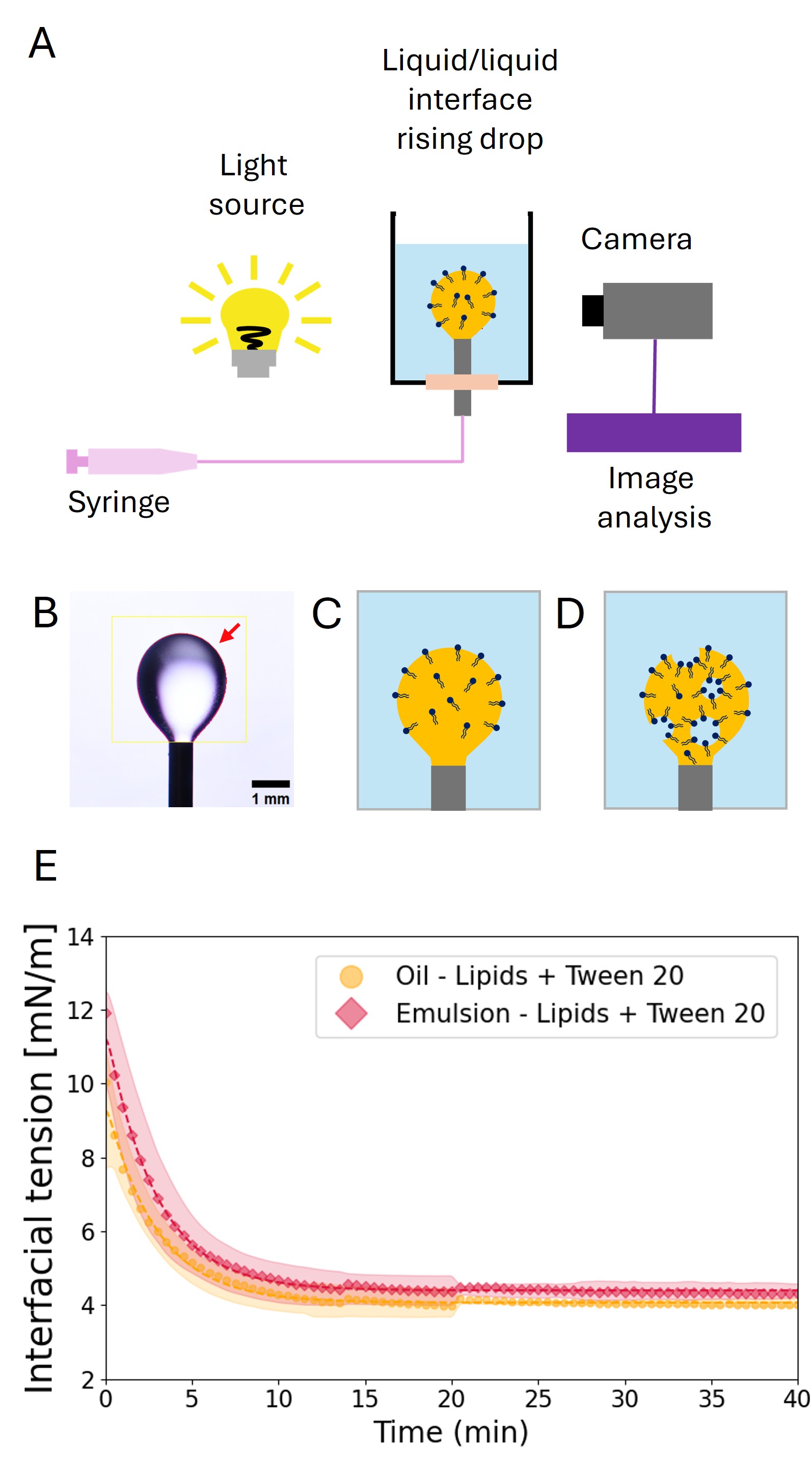}
\caption{Interfacial tension measurements. (A) Schematic illustration of the rising drop tensiometry set-up. (B) Axisymmetric shape profile of the rising drop with the red arrow pointing the final converged guess for the solution of Young-Laplace equation, based on edge detection (see Methods). Illustration of pear-shape like droplet made of a continuous oil phase with surfactants (C) and made of an inverted emulsion (D). (E) Exponential decay of interfacial tension in time. The droplets made of oil (yellow circles) and emulsion (pink diamonds) contain \SI{0.5}{\milli\gram\per\milli\liter} 16:0 Biotinyl PE and \SI{0.25}{\milli\gram\per\milli\liter} Tween 20. Each time point represents a mean value of interfacial tension with the shaded area representing the 95\% confidence intervals (CI). }\label{FIG_5_surfacetension}
\end{figure}

To assess if such values are accessible in our system, we first tested the effect of lipids alone on interfacial tension. To this end, we measured the interfacial tension between an oil containing the lipids and an aqueous phase through a rising drop method (see Fig. \ref{FIG_5_surfacetension} and Methods). Importantly, we also  measured the interfacial tension between an inverted emulsion prepared as mentioned above and a continuous aqueous phase, which matches more accurately the configuration of our system (see Fig. \ref{FIG_5_surfacetension}C-E). {Note that the surfactants require some time to adsorb at the interface of the rising drop. Therefore we track the interfacial tension over time in order to determine its final equilibrium value (see Fig. \ref{FIG_5_surfacetension}E and Methods).}  

\begin{figure}[hbt!]
\centering
\includegraphics[width=0.477\textwidth]{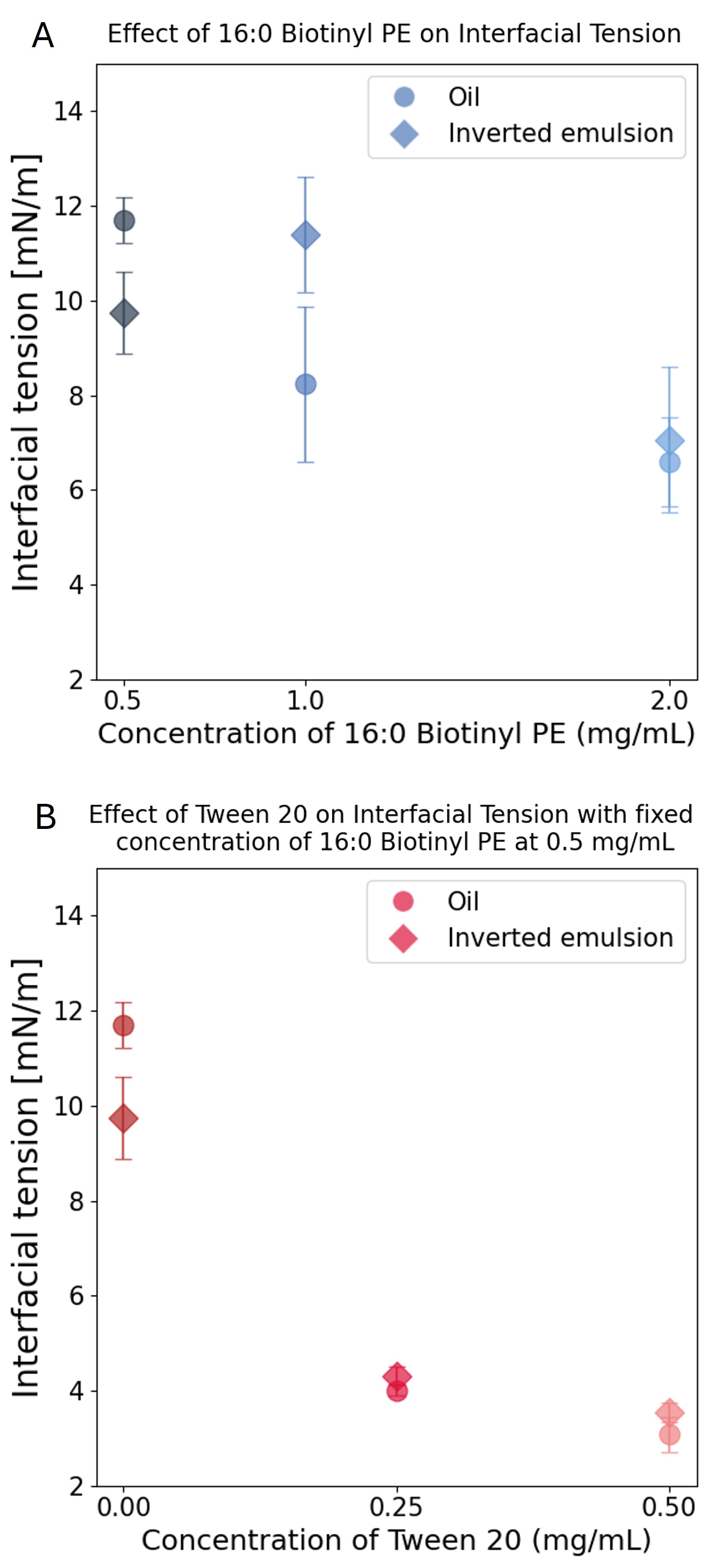}
\caption{Interfacial tension measurements. (A) Interfacial tension as a function of 16:0 Biotinyl PE concentration for oil/water (circles) and inverted emulsion/water (diamonds) interfaces.
(B) Interfacial tension as a function of Tween 20 concentration for oil/water (circles) and inverted emulsion/water (diamonds) interfaces, at a fixed concentration of \SI{0.5}{\milli\gram \per \milli\liter} 16:0 Biotinyl PE lipids.}\label{FIG_6_surfacetension}
\end{figure}

As the concentration of lipids is increased from \SI{0.5}{\milli\gram \per \milli\liter} to \SI{2}{\milli\gram \per \milli\liter} in the oil, the interfacial tension decreases. 
Notably, the same concentrations of lipids lead to similar interfacial tensions with a rising drop made of an inverted emulsion (see Fig. \ref{FIG_6_surfacetension}), thus validating our system for stress measurements \textit{in vivo}. 
However, in both cases, the tension does not go below $\gamma\sim$ \SIrange{6}{7}{\milli\newton\per\meter}, even at \SI{2}{\milli\gram \per \milli\liter} lipids. Yet, at such lipid concentrations we expect the inner water droplets to be quite stable \cite{Pontani2012} and to hinder the targeted self-functionalizing features of the sensor.

To overcome this limitation, the use of Tween 20 as a co-surfactant is critical. Indeed, it is a synthetic surfactant that is largely used to stabilize direct oil in water emulsions, {with a HLB index of 16.7}. Therefore, we expect that the addition of Tween 20 should preferentially lower the oil/water interface at the outer surface of the sensor (positive curvature), while having a more marginal effect on the stability of the inner water droplets whose negative curvature does not favor its insertion.  
To test the effect of Tween 20 on surface tension, we thus measured the oil/water interfacial tension for oils containing a fixed concentration of lipids at \SI{0.5}{\milli\gram \per \milli\liter} and Tween 20 concentrations up to \SI{0.5}{\milli\gram \per \milli\liter} (see Fig. \ref{FIG_6_surfacetension}). 

Strikingly, an addition of \SI{0.25}{\milli\gram \per \milli\liter} Tween 20 already lowers the interfacial tension down to $\sim \SI{4}{\milli\newton\per\meter}$, for both the oil and inverted emulsion systems. Increasing the Tween 20 concentration up to \SI{0.5}{\milli\gram \per \milli\liter} only marginally lowers the tension down to $\sim \SI{3}{\milli\newton\per\meter}$ and could again hinder the destabilization process.
However, in both cases, this value of surface tension is in principle low enough to measure stresses inside soft tissues as discussed above.
In the following, we explore the efficiency of the inverted emulsion destabilization, and the resulting self-functionalization, as a function of surfactant concentrations and composition. 

\subsection{Self-functionalization \textit{in vitro}}

We use agarose to mimic the mechanical properties of soft tissues, and inject oil droplets individually inside the gel in order to track the destabilization of the emulsion through fluorescence imaging (see Methods). 
The injected oil droplets display diameters from 20 to \SI{150}{\micro \meter}, which is compatible with a range of \textit{in vivo} applications.  

\begin{figure}[H]
\centering
\includegraphics[width=0.477\textwidth]{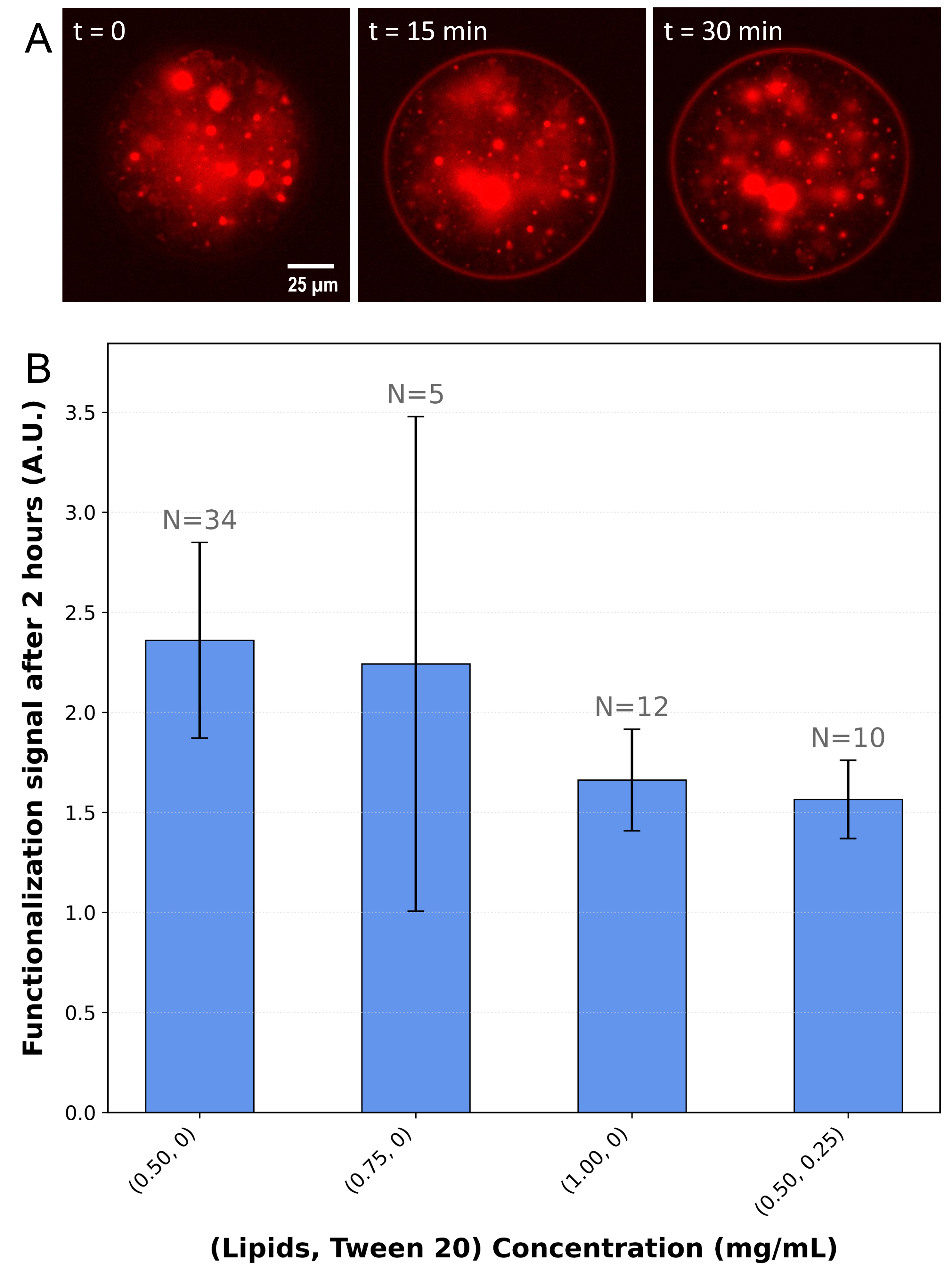}
\caption{
(A) Functionalization of an oil droplet made from an inverted emulsion and injected in agarose over time, with an oil made with \SI{0.5}{\milli\gram\per\milli\liter} of lipids and \SI{0.25}{\milli\gram\per\milli\liter} of Tween 20. At $t=0$, the signal on the oil droplet contour is very low and the functionalization signal yields 1.12. At $t=\SI{15}{\minute}$, a clear ring of fluorescent signal is visible on the outer edge of the oil droplet and the functionalization signal is measured at 1.92, similar to the value of 1.90 obtained at $t=\SI{30}{\minute}$ when the functionalization is stable. (B) Mean values of the functionalization signal measured on the oil droplets \SI{2}{\hour} after injection in the gel, for different concentrations of lipids and Tween 20. Error bars are 95\% confidence intervals of the mean.}\label{functionalization}
\end{figure}

{We label the surface of the inner water droplets by grafting fluorescent streptavidin onto the biotinylated lipids.The fluorescence is initially visible only in the center of the oil droplet, where the inner water droplets are located (see Fig.\ref{functionalization}A, left panel). Over time, a signal appears on the edge of the injected oil droplet (Fig.\ref{functionalization}A, middle and right panels), indicating that some of the inner water droplets have fused with the outer surface over time, allowing the fluorescent streptavidin to relocalize on this outer oil/water interface.}

{We quantify this functionalization process by measuring the fluorescent signal intensity of a \SI{3}{\micro\meter} thick inner ring around the oil droplet, two hours after the injection (see Methods).
On the one hand, we find that a higher concentration of lipids slightly lowers the functionalization signal (see dark to light blue columns in Fig.\ref{functionalization}B). This is due to the fact that a higher lipid concentration tends to stabilize more the inner water droplets.
On the other hand, the addition of Tween 20 at \SI{0.25}{\milli\gram\per\milli\liter} also lowers the functionalization signal (see pink column in Fig.\ref{functionalization}B), making it comparable to the highest lipid concentration (light blue) but with a lower surface tension. We do not explore the functionalization efficiency at higher Tween 20 concentrations, which would decrease its efficiency while only having a marginal effect on surface tension (Fig.\ref{FIG_5_surfacetension}B).}

{Considering the competing requirements of low surface tension ($\gamma \sim$ \SIrange{1}{10}{\milli\newton\per\meter}) and functionalization, we choose \SI{0.5}{\milli\gram\per\milli\liter} lipid, \SI{0.25}{\milli\gram\per\milli\liter} Tween 20 as a formulation that balances well the given constraints.}

\begin{figure}[hbt!]
\centering
\includegraphics[width=0.477\textwidth]{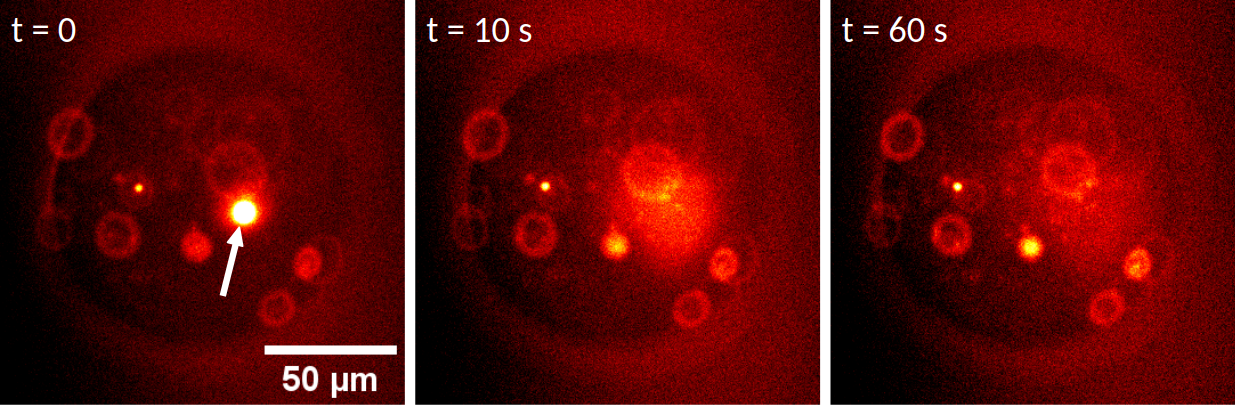}
\caption{Epifluorescence images of an oil droplet loaded with fluorophores and imaged with a 20X objective. The oil contains \SI{0.5}{\milli\gram\per\milli\liter} of 16:0 PE lipids, the inner water droplets contain \SI{25}{\milli\gram\per\milli\liter} Dextran-FITC in \SI{10}{\milli\molar} Tris buffer. Oil droplets are injected in a 3\% agarose gel and imaged over time. The arrow points at an aqueous inner droplet that is still in the oil droplet at $ t = 0 $, but starts releasing its content in the gel at $ t = \SI{10}{\second} $ and, accordingly, we see the growing halo of Dextran-FITC fluorescence as it diffuses away at $t = \SI{60}{\second}$.}\label{figure_dextran}
\end{figure}

{Next, we tested the drug delivery functionality of the oil droplets. 
To do so, we checked if a local release of a loaded agent could be measured with this technique by loading the inner water droplets with large Dextran molecules ($\sim$\SI{150}{\kilo\dalton}) labelled with FITC. 
In this case, because the aqueous phase is made of less expensive and rather robust material, it can be prepared in larger volumes. Therefore the inverted emulsions is made with PVDF filters (see Methods). 
The slow diffusion of this polymeric fluorophore (as compared with FITC alone for instance) allowed us to visualize some release events through which an inner water droplet fuses with the outer surface, thus releasing a halo of fluorescence that diffuses in the surrounding agarose gel (see Fig.\ref{figure_dextran}, left to right panels).}

{In all these tests of functionality, the chosen molecules and fluorescent labels are not surface active agents. Therefore they do not affect the surface tension of the inner water droplets. However, it is important to bear in mind that the encapsulation of surfactants could decrease the self-functionalizing properties of the oil droplet. For instance, the encapsulation of fluorescent colloids did strongly stabilize the inner water droplets oil/water interface, akin to pickering emulsions, and hindered the self-functionalizing properties of the sensors (data not shown).  }

\begin{figure*}
\centering
\includegraphics[width=1\textwidth]{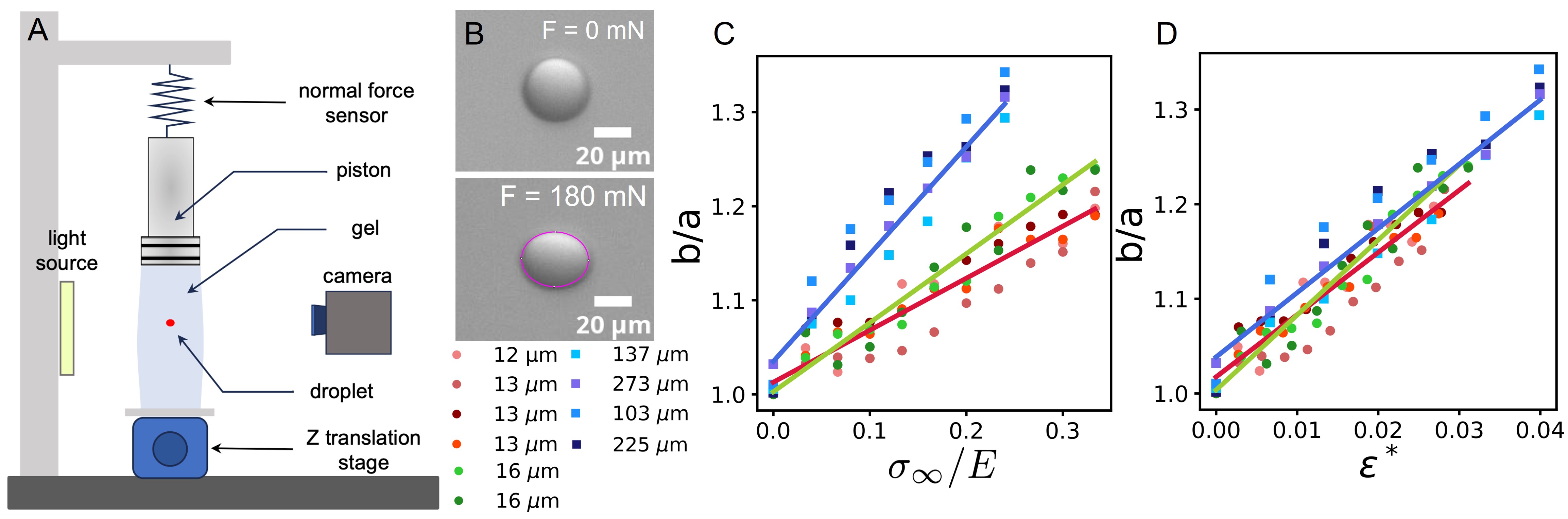}
\caption{ (A) Schematic illustration of mechanical excitation set-up. The system consists of a light source, a cuvette containing an injected droplet in agarose gel, and a camera aligned along the same optical axis. The cuvette is mounted on z-translation stage that enables controlled vertical displacement, allowing a fixed piston to indent the sample. Applied force exerted on the gel is measured using normal force sensor. (B) Brightfield images of soybean oil embedded in 0.6\% agarose gel, illustrating their shape before and after (F=\SI{180}{\milli \newton}) the application of compressive force. Elliptical fit of droplet shape after deformation is shown in magenta. 
(C)Aspect ratio b/a of the oil droplets as a function of the normalized applied stress $\sigma_{\infty}/E$. Legend: radii of the droplets at $\sigma_{\infty} = \SI{0}{\pascal}.$
Squares: inverted emulsions prepared with \SI{0.5}{\milli\gram\per\milli\liter} lipid, \SI{0.25}{\milli\gram\per\milli\liter} Tween 20 and injected in 1\% agarose gel ($E_{1} = \SI{25}{\kilo\pascal}$). Circles: Soybean oil in 0.6 \% agarose ($E_{0.6} = \SI{6}{\kilo\pascal}$).
Red, green, blue lines are linear fits of red, green, and blue data points respectively (droplets with \SIrange{12}{13}{\micro\meter}, \SI{16}{\micro\meter}, and \SIrange{103}{273}{\micro\meter} respectively).
(D) Collapse of all aspect ratio measurements when plotted as a function of the dimensionless parameter $\epsilon^*$ (see equation \ref{epsilon_Dufresne}). 
}\label{graph_manip_ecrasementAgar}
\end{figure*}

\subsection{Mechanical characterization of the sensors}

{Oil droplets have been used inside biological tissues as stress probes~\cite{Campas2014}. In this case, if the oil droplets are embedded in the tissues, the accessible measurement is their local surface deformations.  
In turn, using the Laplace balance on the oil droplet's surface lead to a measurement of the anisotropic stresses at every point of its contour~\cite{Mongera2018}.
Alternatively, these deformations can be approximated in 2D by an elliptical shape, allowing to measure more simply the anisotropic stress from the difference between its principal stresses. 
In this case, the stress anisotropy on the oil droplet is given by the formula $\sigma_{loc} = 2\gamma \left(\mathcal{C}_b - \mathcal{C}_a\right)$, with $\mathcal{C}_b=b/a^2$ and $\mathcal{C}_a=1/2a+a/(2b^2)$ the maximum and minimum curvatures on the ellipsoid, b and a being its long and short semi-axis respectively.
While the Laplace balance is a mechanical equilibrium that holds regardless of constitutive laws of the surrounding medium, if one doesn't know the modulus of the surrounding tissue it is not possible to relate an externally applied force on the tissue to the observed oil droplet deformation. In other words it is used as a local rather than global probe.
However, it can still be a powerful proxy for local stress anisotropies in tissues, as shown in~\cite{Mongera2018}. }

{On the other hand, knowing the elastic modulus of the surrounding medium, one can compute the expected deformation of the oil droplet as a function of an externally applied stress, extending the Eshelby framework to the situation of a liquid inclusion with a surface tension~\cite{Style2015}. 
Therefore, we probed our sensors in a model mechanical excitation set-up~\cite{Tapie2023,Tapie2025} (see Fig.~\ref{graph_manip_ecrasementAgar}A). 
Briefly, oil droplets are injected inside agarose gels mounted inside spectroscopy cuvettes. 
A piston is positioned on top of the gel so that it can apply a known force while the deformation of the oil droplets are visualized by a camera positioned laterally to the cuvette.
At each force step, their deformation is measured by fitting an ellipse to their contour, as shown in Fig.~\ref{graph_manip_ecrasementAgar}B. }

{This approach allowed us to measure the oil droplet aspect ratio and compare it with the theoretical model proposed by Style et al.~\cite{Style2015}, assuming a Poisson’s ratio of $\nu = 1/2$, according to the following relation:}

\begin{equation} \label{Dufresne}
\frac{b}{a} = \frac{1+\frac{10\sigma_{\infty}/E}{6+15\gamma/ER}}{1-\frac{5\sigma_{\infty}/E}{6+15\gamma/ER}},
\end{equation}

{where $\sigma_{\infty}$ is the uniaxial stress applied to the gel, $E$ is the Young’s modulus of the gel, and $\gamma$ is the interfacial tension between the oil droplet and the gel.
Equation \ref{Dufresne} shows that $\sigma_{\infty}$ and $\sigma_{loc}$ can be different, depending on $E$, $R$ and $\gamma$, even if $\sigma_{\infty}$ is the expected stress anisotropy in the gel in the absence of droplet.
In order to probe different regimes, experiments were performed using two agarose gels: 0.6\% agarose with a measured Young’s modulus of $E_{0.6} = \SI{6}{\kilo\pascal}$, and 1\% agarose with $E_{1} = \SI{25}{\kilo\pascal}$.
We injected oil droplets prepared either as inverted emulsions containing \SI{0.5}{\milli\gram\per\milli\liter} lipid and \SI{0.25}{\milli\gram\per\milli\liter} Tween 20, or as pure oil droplets without surfactant.
The latter, associated with a higher surface tension, were used in softer gels in order to probe a regime in which $\gamma/ER \sim 1$. 
In that case, the expected surface tension of the oil droplet is $\gamma = \SI{31}{\milli\newton\per\meter}$~\cite{sahasrabudheDensityViscositySurface2017}~\cite{singhFormulationCharacterizationSoybean2024}. }

{Figure~\ref{graph_manip_ecrasementAgar}C shows the measured aspect ratio of the oil droplets as a function of $\sigma_{\infty}/E$. As predicted by equation~\ref{Dufresne}, large and softer oil droplets embedded in stiff gels for which $R \gg \gamma/E$ (squares in Fig.~\ref{graph_manip_ecrasementAgar}C) exhibit an aspect ratio independent of their size (note the large spread of sizes, ranging from $\sim 100$--$\SI{300}{\micro\meter}$). 
Conversely, when $\gamma/ER \sim 1$ (circles in Fig.~\ref{graph_manip_ecrasementAgar}A), surface tension resists deformation, resulting in a smaller aspect ratio compared to that of an infinitely deformable oil droplet.}

{From equation~\ref{Dufresne}, the aspect ratio of the oil droplet is expected to depend only on the dimensionless parameter
\begin{equation} \label{epsilon_Dufresne}
\epsilon^* = \frac{\sigma_{\infty}/E}{6+15\gamma/ER}
\end{equation}
that contains information on its size, surface tension and gel stiffness.
As shown in Fig.~\ref{graph_manip_ecrasementAgar}D, all data collapse onto a single curve when plotted as a function of $\epsilon^*$.
This analysis confirms the importance of surrounding material properties, together with the size of the oil droplets and their surface tension, to understand their deformation inside a material subject to mechanical load.
In other words, if the material properties of the surrounding tissue and sensor are known, the global stress applied on the tissue can be derived through the above-described framework. If only the sensor properties are known, its use is limited to probing local anisotropic stresses on the oil droplet.}

\subsection {Injection inside biological material}

\begin{figure*}
\centering
\includegraphics[width=1\textwidth]{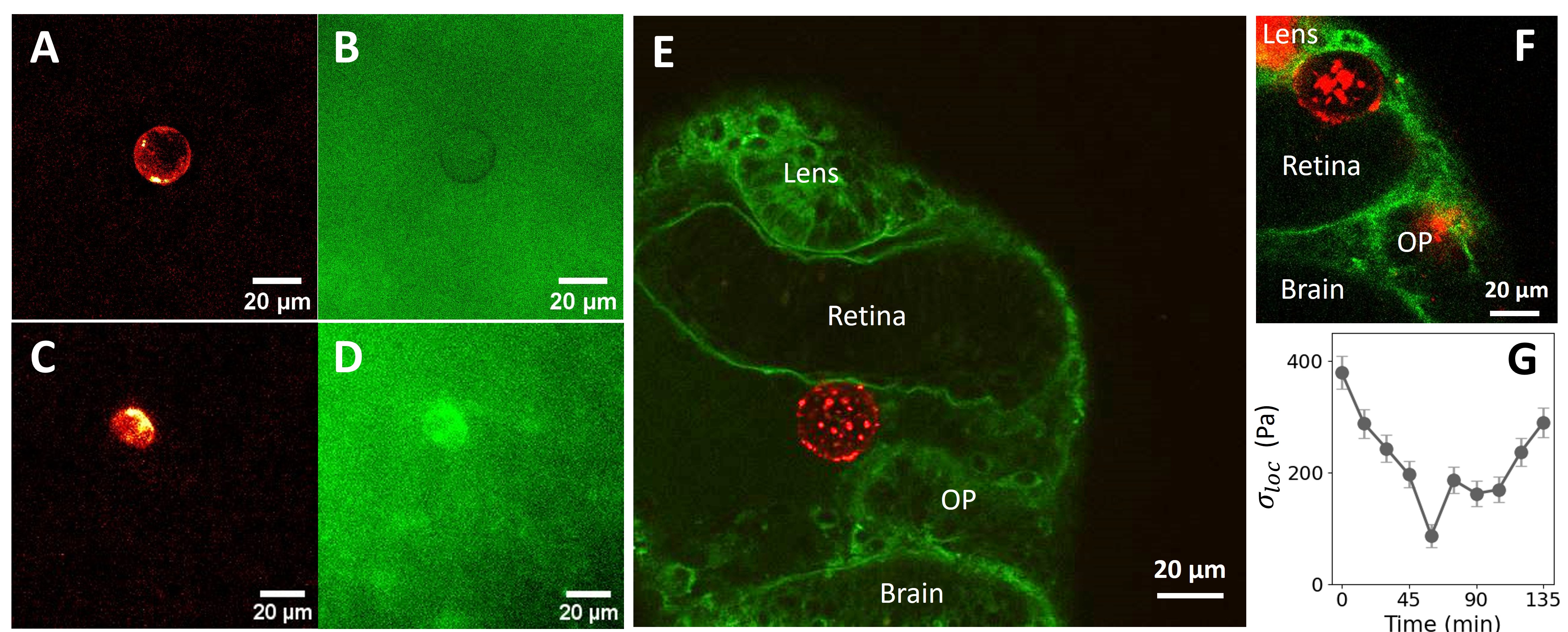}
\caption{
(A-D) Self-functionalizing oil droplet injected in a forebrain organoid. (A) Epifluorescence images of a self-functionalizing oil droplet labelled with fluorescent streptavidin (red channel) and injected in an undeformed organoid whose cells are labelled in green as seen in (B). (C-D) Laser-sheet microscopy image of a deformed oil droplet (C) inside the same organoid (D) after it was deformed by being sucked into a capillary. (E-F) Confocal microscopy images of self-functionalized oil droplets prepared with \SI{0.625}{\milli\gram\per\milli\liter} lipid, injected in TgBAC(lamC1:lamC1-sfGFP) transgenic embryos, in which Laminin-rich basement membranes are visible in green. The inner water droplets are labelled with fluorescent streptavidin (red channel) and the sensors are injected near the Retina-OP interface (E) or within the eye at the Lens-Retina interface (F). Orientation in (E) and (F): anterior to the right, lateral to the top. (G) Evolution of the local anisotropic stress ($\sigma_{loc}$) measured for the droplet presented in panel (F) over time, calculated from its ellipticity extracted from brightfield images (uncertainty estimated from  the  image  resolution). Interfacial tension corresponding to \SI{0.625}{\milli\gram\per\milli\liter} lipids was estimated to be $10.61 \pm \SI{0.71}{\milli\newton\per\meter}$ by linear extrapolation of the result presented in~\ref{FIG_6_surfacetension}A.}\label{injection_invivo}
\end{figure*}

As a proof of concept, we tested the oil droplets in two different biological systems. First, we injected inverted emulsion oil droplets labeled with fluorescent streptavidin into dorsal forebrain organoids (see Methods). 
The organoids, that were embedded in agarose, were first imaged immediately after injection, showing fluorescent spherical oil droplets distributed within the tissue (Fig. \ref{injection_invivo}A-B). A piece of agarose containing the organoid was then cut out and placed into a circular capillary. 
This procedure allowed to deform the organoid, which was further imaged through light sheet microscopy (see Methods). Elliptical oil droplets were clearly visible at approximately \SI{50}{\micro\meter} deep within the tissue (Fig. \ref{injection_invivo}C-D), demonstrating that they have sufficiently low interfacial tension for \SI{20}{\micro\meter} oil droplets to be deform under mechanical stress within soft tissues such as organoids \cite{ryuTransparentCompliant3D2021}.
Since in this case the absolute modulus of the tissue is unknown, we compute the local anisotropic stress on the oil droplet through the formula $\sigma_{loc} = 2\gamma \left(\mathcal{C}_b - \mathcal{C}_a\right)$, as described in the previous section. 
For the oil droplet shown in Fig.\ref{injection_invivo}C, we can thus estimate an anisotropic stress on the droplet $\sigma_{loc} \simeq \SI{490 +- 80}{\pascal} $ {(uncertainty calculated from the image resolution). This value compares well with previously reported values measured with solid sensors in deformed cellular aggregates~\cite{Souchaud2022}.}

Second, injections inside zebrafish embryos were performed at an early developmental stage (14-somites). The targeted area is near the eye and olfactory placode (OP), which is close to the surface and allows easy imaging. As shown in Fig. \ref{injection_invivo}E, the oil droplets are well inserted in the embryos and their signal allows for their surface tracking \textit{in situ}. 
{Since this area of the embryo is mostly composed of extracellular matrix and loosely-arranged migrating neural crest cells as well as other cell types of the peri-occular mesenchyme~\cite{Bryan2020}, the oil droplet does not exhibit any measurable deformation. Conversely, when the sensors are injected in between two compact tissues such as the retina and the lens, as in Fig.~\ref{injection_invivo}F, their deformation is significant and measurable. Furthermore, the deformation of the oil droplet can be tracked over time, leading to an estimation of the local anisotropic stress evolution on the droplet during development in this particular area of the zebrafish embryo (see Fig.\ref{injection_invivo}G). The decrease of local anisotropic stress could be assigned to either relaxation of local tissue stresses through viscous dissipation or to an increase of the stress orthogonal to the long axis of the droplet.}

\section{Methods}\label{sec11}

\textbf{Oil and aqueous phases composition} --
The continuous oil phase consisted of soybean oil (Sigma-Aldrich) containing 16:0 Biotinyl PE phospholipids (Avanti Polar Lipids) and Tween 20 (Merck) at various concentrations. Briefly, a stock solution of phospholipids dissolved in chloroform was transferred to a glass vial at the desired concentration. The solvent was then evaporated under a nitrogen stream, after which \SI{4}{\milli\liter} of soybean oil containing Tween 20 was added to the dried lipids. The mixture was sonicated for 30 minutes at room temperature.
When Tween 20 was used, it was directly dissolved in soybean oil at final concentrations of \SI{0.25}{\milli\gram\per\milli\liter} or \SI{0.50}{\milli\gram\per\milli\liter}.

\SI{50}{\micro\liter} of aqueous phase is then prepared for the formation of the inverted emulsion. For self-functionalizing experiments, it contains \SI{0.2}{\milli\gram\per\milli\liter} Streptavidin, Alexa Fluor 546 conjugate (Thermo Fischer) in 10 mM Trizma hydrochloride buffer (Tris), pH = 7.4 (Sigma Aldrich). 
For drug release experiments, the water phase is made of \SI{25}{\milli\gram\per\milli\liter} FITC Dextran (FD 150, COGER) in 10 mM Trizma hydrochloride buffer (Tris), pH = 7.4 (Sigma Aldrich).

\textbf{Water-in-oil (W/O) emulsion preparation} --
Two \SI{1}{\milli\liter} luer-lock syringes (Fisher scientific) were filled with \SI{500}{\micro\liter} (syringe A) and \SI{300}{\micro\liter} (syringe B) of freshly vortexed surfactant-containing oil. 
After removal of any air bubble, syringe A was connected to \SI{5}{\micro\meter} pore size hydrophobic SPG pumping membrane connector (SPG Technology) and the oil was pushed through the connector. 
A 18 G needle (Terumo) was mounted on syringe B. 
About \SI{20}{\micro\liter} of oil was coarsely emulsified in the \SI{50}{\micro\liter} of aqueous phase through back and forth movements (5 to 10 times) inside the needle of syringe B. 
This coarse emulsion was then aspirated into syringe B before it was mounted onto the other side of the SPG membrane connector, while still making sure that no air bubbles are trapped in the syringe.
The water droplet of coarse emulsion was then pushed back and forth through the connector, going from one syringe to the other. A total of five passages through the SPG membrane were completed, leading to the inverted emulsion used in the next stages.
After each emulsion preparation, the SPG membrane connector was cleaned by flushing it with \SI{1}{\milli\liter} of Mili-Q water, followed by \SI{1}{\milli\liter} of soybean oil containing \SI{0.125}{\milli\gram\per\milli\liter} 16:0 Biotinyl PE. 

When the material constituting  the  inner  water  droplets  is  not fragile biological material (Dextran, Fluorescein...), the emulsification method can be more simply performed in two steps. 
First, a coarse emulsion is made by mixing the oil (\SI{800}{\micro\liter}) and the water phase (\SI{100}{\micro\liter}) going through the needle several times. 
Then this coarse emulsion is pushed once through a PVDF filter with a pore size of \SI{5}{\micro\meter} (Millex, ref. SLSV025LS), and about \SI{400}{\micro\liter} of emulsion is recovered, due to the large dead volume in the filter.

{In both cases (SPG membrane or filter), this emulsification step can be completed in a few minutes. The prepared emulsion should then be used immediately for injection as detailed below.}

\textbf{Inner water droplets imaging} --
After preparing the emulsion with the protocols detailed above, a few \SI{}{\micro \litre} are inserted between a glass slide and a coverslip. The emulsion sample is then imaged through spinning disk confocal microscopy with a 60X oil immersion objective (1.4 N.A.) (see typical image in Fig. \ref{figure_droplet_size}A). 

\textbf{Interfacial tension measurements} --
Interfacial tension between oil and water were measured by rising drop tensiometry. A blunt needle (G 23, PHYMEP) was inserted through the bottom of a polystyrene macro cuvette (Fisher scientific), filled with an aqueous solution of 10mM Tris (pH = 7.4 ). The oil phase (or inverted emulsion) was injected through the needle until a rising pear shaped oil droplet was formed (see \ref{FIG_5_surfacetension}A-D). The set-up was illuminated in transmission thanks to a led panel and the oil droplet was imaged with a camera over time (an image every 5 minutes during 20 to 40 minutes).  

The interfacial tension was estimated from the axisymmetric shape profile of the rising drop using \textit{Pendant Drop} plugin in Fiji (see \ref{FIG_5_surfacetension}B) \cite{Daerr2016}.

From each condition, we get the mean measurement of surface tension at each time point, which we fit with an exponential decay to a finite value function (see \ref{FIG_5_surfacetension}E). 
This finite value obtained from the fit is our reported equilibrium value in Fig \ref{FIG_6_surfacetension}A-B .

\textbf{Microinjection in agarose} -- The microinjection of inverted emulsion oil droplets was performed under a binocular microscope (Olympus, SZX2-FOFH). Borosilicate standard wall without filament capillaries of 0.69 mm inner and 1.2 mm outer diameter (Harvard Apparatus, 30-0042) were pulled using a micropipette puller (P-97, Sutter Instrument) under optimized settings (pressure = 500, heat=600, pull=120, velocity = 50, time = 225).
To produce long, fine and initially sealed tips. Approximately \SI{15}{\micro\liter} of freshly prepared inverted emulsion was loaded into the capillary using \SI{20}{\micro\liter} microloader pipette tips (Eppendorf). 
The capillary tip was then manually cut using scalpel, achieving openings of a few micrometers in diameter. 
The capillary was finally mounted onto a holder connected to the pressure control system (Femto Jet 4i, Eppendorf). 
{To ensure constant inverted emulsion volume injection, two approaches can be used: either the glass capillaries should be pre-opened with the reproducible tip opening, followed by standardized microinjector parameters; or if the capillary tips are less controlled in size, the microinjector can be calibrated in advance for each capillary, in order to deliver the desired volume. Note that cheaper manual injectors can also be used (such as Eppendorf celltram oil pumps) but require a lot more practice by the users in order to reach precise volumes.}

A solution of 0.5\% (w/v) low-melting agarose (Sigma Aldrich) gel, prepared in 10 mM Tris buffer was poured in 35 mm glass bottom dish (Ibidi). 
The injection was performed at an intermediary stage, before full gelling was achieved. 
The microcapillary tip was positioned as close as possible to the bottom of the dish, in order to remain within the working distance of the inverted microscope used for observation of self-functionalization dynamics.
After injection, 1 mL of 10 mM Tris buffer were gently added on top of the gel to avoid evaporation effects and the dish was additionally sealed with its lid and wrapped with parafilm. 

{\textbf{Mechanical characterization of stress sensor} --
To probe the stress-sensing properties of the oil droplets, we used a mechanical excitation set-up, described in detail in ref.~\cite{Tapie2023,Tapie2025}, with slight modifications. Briefly, a gel slab containing the oil droplets can be lifted vertically using a manual micrometric Z stage to indent a fixed piston, itself mounted on a force measurement setup. The force sensor consists in a dual cantilever beam (whose stiffness has been independently measured). The deflection of the cantilever is measured with a capacitive sensor, from which the normal $F_n$ is deduced (range \SIrange{0}{2}{\newton}), with a measurement noise of about \SI{10}{\milli\newton}. 
In order to work with a cube of agarose gel the bottom of the cuvette was filled with \SI{3}{\milli\liter} polydimethylsiloxane (PDMS). Moreover, two thin spacers made of Plexiglas (thickness \SI{1}{\milli\meter}) can be placed on two opposing sides of the cuvette, enabiling, after their removal, lateral expansion of the gel during compression. 
Agarose gels at concentrations 1\% and 0,6\% were used to tune the corresponding elastocapillary length.
Freshly prepared inverted emulsion was injected in 1\% agarose gel, as described above, before full gelling was achieved. 
The applied force was increased in \SI{0.1}{\newton} steps, starting from 0 N up to maximum of \SI{0.6}{\newton}, while monitoring the shape of oil droplets with a Pointgrey camera (BFLY-U3-23S6C-C) facing a side of the cuvette. Once the maximal force was reached, the cuvette was shifted downward to reduce the load to \SI{0}{\newton}, allowing oil droplets to relax and go back to their spherical shape. For the 0.6\% agarose gel, soybean oil droplets were injected instead, and the applied force was increased in \SI{0.02}{\newton} steps from \SI{0}{\newton} up to \SI{0.2}{\newton}, following the same protocol.}

\textbf{Microinjection in zebrafish embryos} --
Using the same capillaries and injection protocol, oil droplets of inverted emulsion were injected in zebrafish embryos mounted in 0.5\% low melting agarse at 14-somite stage (corresponding to 16 hpf). {It should be noted that depending on the tissue accessibility and mechanical properties, glass capillaries used for injection can be tailored by adjusting the micropipette puller settings. For example, if the target site is located superficially underneath the stiff structure, such as skin, shorter capillaries are preferable as they are less prone to bending and provide better penetration. In contrast, if the target is located deeper, within softer tissue, usually longer and thinner capillaries are required, in order to reach the site effectively and not cause damage.} 
The embryos are obtained from a transgenic zebrafish line TgBAC(lamC1:lamC1-sfGFP) expressing Laminin gamma1 fused to sfGFP, under the control of its own regulatory sequences, as described in \cite{Yamaguchi2022}.

All the zebrafish experiments were made in agreement with the European Directive 210/63/EU on the protection of animals used for scientific purposes, and the French application decree "Décret 2013–118". 
The fish facility has been approved by the French "Service for animal protection and health", with the approval number B-75-05-25. 
The DAP number of the project is APAFIS \#47427-2023121817395554 v6. 

\textbf{Microinjection in organoids} --
Oil droplets were injected into a dorsal forebrain organoid derived from human induced pluripotent stem cells (hiPSCs). The organoid was generated and maintained in long-term culture following the protocol described in~\cite{Sloan2018}, with adaptations from~\cite{Yoon2019}. For quality control, routine mycoplasma testing, SNP analysis, and immunohistochemistry were performed to ensure genomic integrity and correct brain region specification. The hiPSC line used was the UKBi013-A-GCaMP6f iPSC reporter line, derived from the parental line UKBi013-A (https://hpscreg.eu/cell-line/UKBi013-A) via CRISPR/Cas9-mediated genome editing, as described in detail in~\cite{Li2025}. At the time of injection, the organoid was 100 days old.
Samples were embedded in 1.5\% low-melting-point agarose (Invitrogen) prepared in sample medium and loaded into glass capillaries (Drummond, Wiretrol II). 
The capillary was then placed into the imaging chamber, which was filled with sample medium. Prior to imaging, the sample was gently extruded from the capillary to ensure direct exposure to the light-sheet.

\textbf{Imaging} --
The oil in water emulsion was imaged through spinning disk confocal microscopy (Spinning Disk Xlight V2, Gataca systems) {using a 60X oil immersion objective}.
Oil droplets in agarose and underformed organoids were imaged through epifluorescence microscopy (inverted Nikon Eclipse Ti2-E), using a Nikon, PLAN APO Lambda D, 20x/0.8 objective. 
In order to track the oil droplets over time, z-stacks were acquired to compensate for drifts. 
Image analysis was then performed in the identified equatorial plane of each oil droplet.

Light-sheet fluorescence imaging was performed using a previously described setup~\cite{hubertVersatileOpenSource2025}. Two continuous-wave lasers were used for excitation: a 561 nm laser (Vortran) for imaging red fluorescence and a 488 nm laser (Oxxius) for imaging green fluorescence. The light-sheet was generated via a galvanometric mirror (Thorlabs, GVS001) and directed into the sample through an illumination objective (Zeiss, 5X A-Plan, NA=0.12). Fluorescence emission was collected using a detection objective (Leica,  HC FLUOTAR L 25x/0,95W VISIR) and a tube lens (Olympus), resulting in a total system magnification of 22.5X. An appropriate emission filter was placed at the detection path depending on the fluorescence channel being recorded. Images were acquired using a Hamamatsu camera. The total illumination power delivered to the sample was kept below 1 mW.

Zebrafish embryos were imaged through Leica TCS SP8 MP upright microscope, using 25X 0.95 NA Water dipping objective.

\textbf{Image analysis} --
{The size distribution of the inner water droplets (as shown in Fig.\ref{figure_droplet_size}) was measured on confocal images acquired with a 60X oil immersion objective and with parameters that prevented any saturation. First, a manual intensity threshold was applied to ensure the detection of all visible water droplets.}
Then, a watershed function was performed to separate the few water droplets that were too clustered and could appear merged by the thresholding step.
Last, the shape and size of all particles detected this way was measured, removing all particles that are smaller than {6 square pixels, corresponding to a diameter of $\sim$\SI{200}{\nano\meter},} or that have a circularity lower than 0.5 to remove potential noise.
In practice, less than 1\% of particles detected were in these criteria, which indicates that the signal and detection algorithm was accurate.
All diameters measured below the microscope resolution (taken as \SI{300}{\nano \meter}) were assigned the value d=\SI{300}{\nano \meter}.
These water droplets accounted for less than 5\% of the detected ones and thus have a negligible effect on the mean size and standard deviation.

Oil droplet functionalization analysis (Fig.\ref{functionalization}) was done in three major steps.
First, the images were prepared for contour detection with saturation and bleaching compensation.
The bleaching compensation is applied first by running a histogram-matching algorithm on series of the same field of view over time.
Then, the saturation is done by setting all pixels that have an intensity higher than 10 times the intensity of the background. From experience, this setup ensured that no pixel of interest (from the oil droplet contour) were saturated, but most pixels coming from the very bright inner water droplets were saturated.
Second, the oil droplet contour was detected by hand, pointing at 3 pixels on its edge. For very bright contours (which means a very high functionalization), an automatic detection algorithm was used and it resulted in no difference with the manually detected contour.
Last, the normalized intensity of each oil droplet (which is the final proxy for functionalization) is calculated by measuring the mean pixel intensity in a \SI{3}{\micro\meter} ring around its contour, and dividing this value by the mean background intensity.

\section{Discussion}\label{sec12}

This article presents a method to design custom biocompatible force sensors with added capabilities, either regarding functionalization or local drug delivery. The method is based on an inverted emulsion for which very little material is needed. This inverted emulsion can be produced with commercial SPG membrane connectors or even classical filters. In both cases, the emulsion synthesis itself does not require any specific equipment such as microfluidics, syringe pumps or large emulsification apparatus. In fact, it can be done in a few minutes in a minimalistic lab environment, making this method available to a large range of research groups.

After optimizing the synthesis, we show that the formulation of the emulsion can be tuned to combine different properties. First, the use of soybean oil and phospholipids make these soft matter tools inherently biocompatible. 
Second, the use of Tween 20 as a co-surfactant efficiently lowers the surface tension of the oil droplets, making them soft enough for stress sensing applications at cellular scales in a range of biological tissues, as evidenced by our injections inside organoids. These experiments ultimately allowed us to quantify a local stress of a few hundreds of pascals upon external deformation of the organoid. 
Third, this formulation still ensures the progressive destabilization of the inverted emulsion, which allows for self-functionalization and drug delivery properties of the sensors. 
The destabilization is not necessarily complete, as shown by the remaining inner water droplets visible in the injected oil droplet in the zebrafish embryos. However, using fluorescent binders or molecules still allows one to track precisely the extent of surface functionalization and the repartition of the binders in a given tissue. 
Moreover, these remaining inner water droplets may be turned into an advantage. Indeed, the presence of discrete patches in the oil droplets or on their surface can be used as a referential for posture measurements. This could for instance allow to track events such as rotations inside biological tissues.

In addition to this, the self-functionalization of the surface through phospholipids offers a wide platform for functionalization. Here we used biotinylated lipids that can be decorated with streptavidin and further grafted with any biotinylated molecule. One can also use phospholipids carrying a NTA-Ni group allow for the grafting of poly-histidine tagged molecules \cite{Pontani2016a} or even phospholipids that offer a click chemistry platform on their hydrophilic head. 

{Importantly, these sensors could allow to decipher biological processes through different approaches. First, comparing sensors that interact with their surroundings through adhesive or passive contacts could help discriminate between different stress sources. For instance, whether a tissue is locally compressed laterally or pulled axially can lead to the same extension. Adhesive sensors could not only measure the stresses associated to extension but inform on its origin. Additionally, the self-functionalizing sensors can be used through a perturbative approach. The release of local drugs of cellular activity could thus unravel the contribution of cell generated forces into specific processes. 
Finally, thanks to the general framework that we introduced regarding their mechanical characterization, one can rationally measure either local stresses applied on the sensors, or unravel global stresses applied on a material whose properties are know.
Therefore, this simple, yet versatile, soft matter tool could therefore be used for a range of applications inside biological materials. }

\backmatter



\bmhead{Acknowledgements}

The authors thank Marco Ribezzi and Victor Zahoransky for their help with the use of the SPG membrane connectors, Jerome Fresnais for his help with surface tension measurements, and Loren Jorgensen for her help with the agarose gel rheometry. The authors also thank France Lam and Chloe Chaumeton for their support at the IBPS imaging facility. Finally, the authors thank Raphael Voituriez for fruitful discussions regarding mechanical modelling. Marie Anne Breau and Lea-Laetitia Pontani acknowledge financial support from Agence Nationale de la Recherche (MECAMATRIX, ANR-23-CE13-0025). Olga Vasiljevic aknowledges financial support from the i-Bio international PhD Program.

\section*{Declarations}

\bmhead{Competing interests}
There are no competing interests to declare.

\bmhead{Data availability}
Data will be available in a public repository.

\bmhead{Author contributions}
LLP, MB and JF conceived the study. OV, NH and LLP designed the experiments. OV, NH, EW, AH, LK and CH performed experiments. OV and NH analyzed data. OV, NH, VB, JF, EW, MB and LLP interpreted the results. OV, NH and LLP wrote the manuscript with feedback from all authors. LLP supervised the project.

\printbibliography

\end{document}